\documentclass[preprint2]{aastex}
\usepackage{psfig}

\slugcomment{submitted to ApJ}

\shorttitle{X-Ray Jets of 4U 1755--33}
\shortauthors{Kaaret et al.}

\begin{document}

\title{Evolution of the X-Ray Jets from 4U 1755--33}

\author{P.\ Kaaret\altaffilmark{1}, S.\ Corbel\altaffilmark{2}, J.A.\
Tomsick\altaffilmark{3}, J.\ Lazendic\altaffilmark{4}, A.K.\
Tzioumis\altaffilmark{5}, Y.\ Butt\altaffilmark{6}, R.\
Wijnands\altaffilmark{7}}

\altaffiltext{1}{Department of Physics and Astronomy, University of
Iowa,  Van Allen Hall, Iowa City, IA 52242, USA;
philip-kaaret@uiowa.edu}

\altaffiltext{2}{AIM - Unit\'e Mixte de Recherche CEA - CNRS -
Universit\'e Paris VII - UMR 7158, CEA Saclay, Service d'Astrophysique,
F-91191 Gif sur Yvette, France}

\altaffiltext{3}{Center for Astrophysics and Space Sciences, University
of California at San Diego, La Jolla, CA 92093-0424, USA}

\altaffiltext{4}{Kavli Institute for Astrophysics and Space Research,
Massachusetts Institute of Technology, 77 Massachusetts Avenue,
Cambridge, MA 02139}

\altaffiltext{5}{Australia Telescope National Facility, CSIRO, P.O.\
Box 76, Epping, NSW 1710, Australia}

\altaffiltext{6}{Harvard-Smithsonian Center for Astrophysics, 60 Garden
St., Cambridge, MA 02138, USA}

\altaffiltext{7}{Astronomical Institute `Anton Pannekoek', University
of Amsterdam, Kruislaan 403, 1098 SJ Amsterdam, The Netherlands}

\begin{abstract}

We report on new X-ray observations of the large-scale jets recently
discovered in X-rays from the black hole candidate 4U 1755-33.  Our
observations in 2004 show that the jets found in 2001 are still present
in X-rays.  However, sensitive radio observations in 2004 failed to
detect the jets.  We suggest that synchrotron radiation is a viable
emission mechanism for the jets and that thermal bremsstrahlung and
inverse-Compton emission are unlikely on energetic grounds.  In the
synchrotron interpretation, the production of X-rays requires
acceleration of electrons up to $\sim 60$ TeV, the jet power is $\sim 4
\times 10^{35} \rm \, erg \, s^{-1}$, and the radio non-detection
requires a spectral index $\alpha > -0.65$ ($S_{\nu} \propto
\nu^{\alpha}$) which is similar to the indexes found in lobes
surrounding some other compact objects.  We find an upper limit on the
flux of 4U 1755-33 in quiescence of $5 \times 10^{-16} \rm \, erg \,
cm^{-2} \, s^{-1}$ in the 0.3-8~keV band.

\end{abstract}

\keywords{black hole physics: general -- stars: black holes -- stars:
individual (4U 1755-33) -- stars: winds, outflows -- X-rays: binaries}

\section{Introduction}

Relativistic jets are often produced by accreting compact objects,
including both stellar-mass X-ray binaries and active galactic nuclei.
Jets appear to be important in the dynamics of the overall accretion
flow in such systems and a substantial fraction of the accretion energy
of X-ray binaries may be dissipated in jets.  Understanding the
properties and production of jets is important for our understanding of
the energetics and dynamics of the accretion process.  Further, the
jets from X-ray binaries in the Milky Way are a potentially significant
source of energy input to the Galactic ISM and, if hadronic, a likely
source of a cosmic rays \citep{heinz02} and a possible source of
Galactic light element nucleosynthesis \citep{butt03}.


Recent XMM-Newton observations have led to the discovery of a large
scale X-ray jet from the long-term X-ray transient and black hole
candidate 4U 1755--33 \citep{angelini03}.  The X-ray source 4U 1755--33
was discovered with Uhuru \citep{jones77} and was, therefore, active in
1970 and may have been active earlier.  It was later found to have an
unusually soft spectrum \citep{whitem84,white84} and a hard X-ray tail
\citep{pan95} suggesting a black hole candidate.  The source shows
X-rays dips which indicate that the system has a high inclination and
an orbital period of 4.4 days \citep{white84,mason85}.  The source was
still active in 1993 \citep{church97}, and then found in quiescence in
1996 \citep{roberts96}.  The source was active for at least 23 years.
The distance to the source is poorly constrained.  It is likely greater
than 4~kpc because the optical counterpart (identified during outburst;
McClintock, Canizares, \& Hiltner 1978) was not detected in quiescence
\citep{wachter98}, but less than 9~kpc because of the low level of
visual extinction \citep{mason85}.

\citet{angelini03} found a linear X-ray structure which is roughly
symmetric about the position of 4U 1755--33 extending about $3\arcmin$
to the north-west and $3\arcmin$ to the south-east of the black hole
candidate.  There appear to be multiple knots in the jets.  For
estimated distances of 4--9~kpc, the angular size corresponds to jet
lengths of 3--8~pc. Therefore, the jet must have taken at least 10-30
years to form.  The source was active for at least 23 years which
appears sufficient to have formed the jet.  Chandra observations do not
show point sources along the jet, but do give a detection of emission
over the area of the jet \citep{park05}.  This indicates that the jet
emission seen with XMM-Newton is truly diffuse and not an alignment of
point sources.

The primary questions concerning the jet are: what is the X-ray
emission mechanism, how is the jet powered, and what is the total
energy in the jet?  We obtained new observations of 4U 1755--33 using
XMM-Newton and the Australia Telescope Compact Array (ATCA) to attempt
to measure a multiwavelength spectrum and observe the evolution of the
jet.  We describe the X-ray observations and analysis in \S 2, the
radio observations and analysis in \S 3, and draw conclusions regarding
the properties of the jet in \S 4.

\begin{deluxetable}{rlllrrl}
\tabletypesize{\scriptsize}
\tablecaption{XMM-Newton X-ray sources near 4U 1755-33
  \label{xmmsources}}
\tablewidth{0pt}
\tablehead{ &
  \colhead{RA} & \colhead{DEC} & \colhead{Error} & 
  \colhead{Flux A}  &
  \colhead{Flux B}  &
  \colhead{Counterparts} }
\startdata
  1 & 17 59 00.86 & -33 45 48.1 & 0.2 & 243.9 $\pm$10.0 & 171.9 $\pm$ 4.0 & AW2003 1, 2mass \\
  2 & 17 57 58.88 & -33 46 24.9 & 0.4 & 218.4 $\pm$12.1 &  45.1 $\pm$ 2.7 & 2mass \\
  3 & 17 59 21.83 & -33 53 15.6 & 0.7 &  77.1 $\pm$ 7.1 &  14.9 $\pm$ 1.9 & AW2003 3 \\
  4 & 17 58 26.89 & -33 59 53.8 & 0.8 &  76.2 $\pm$ 9.7 &    -            & \\
  5 & 17 58 42.02 & -33 41 48.8 & 0.3 &  29.3 $\pm$ 3.8 &  71.6 $\pm$ 2.9 & 1WGA J1758.6-3341 \\
  6 & 17 59 22.79 & -33 50 25.2 & 0.7 &  63.8 $\pm$ 7.0 &  24.4 $\pm$ 2.3 & 1WGA J1759.3-3350, 2mass \\
  7 & 17 58 20.25 & -33 42 51.8 & 0.6 &  63.4 $\pm$ 7.7 &  33.7 $\pm$ 2.2 & \\
  8 & 17 58 49.48 & -33 41 40.8 & 0.5 &  56.6 $\pm$ 5.7 &  42.7 $\pm$ 2.7 & 2mass \\
  9 & 17 58 36.67 & -33 40 27.3 & 0.6 &    -            &  55.0 $\pm$ 5.7 & \\
 10 & 17 58 38.88 & -33 57 00.3 & 0.4 &    -            &  49.0 $\pm$ 2.5 & \\
 11 & 17 58 21.03 & -33 46 54.7 & 0.3 &  34.1 $\pm$ 4.3 &  41.8 $\pm$ 1.9 & 2mass \\
 12 & 17 59 27.97 & -33 49 10.3 & 1.5 &  37.0 $\pm$ 4.5 &    -            & \\
 13 & 17 58 44.04 & -33 46 13.0 & 0.5 &  29.2 $\pm$ 3.4 &  11.7 $\pm$ 1.2 & \\
 14 & 17 57 40.65 & -33 45 53.9 & 0.6 &    -            &  25.4 $\pm$ 2.3 & \\
 15 & 17 58 59.19 & -33 50 16.3 & 0.5 &  25.3 $\pm$ 3.2 &  20.4 $\pm$ 1.5 & \\
 16 & 17 58 44.16 & -33 45 09.5 & 0.6 &  24.5 $\pm$ 3.2 &  12.2 $\pm$ 1.4 & \\
 17 & 17 59 03.24 & -33 59 18.1 & 0.9 &    -            &  17.8 $\pm$ 2.2 & \\
 18 & 17 58 51.23 & -33 49 13.8 & 0.7 &    -            &  17.1 $\pm$ 1.4 & \\
 19 & 17 58 25.56 & -33 57 46.8 & 2.8 &    -            &  15.8 $\pm$ 1.7 & \\
 20 & 17 58 24.23 & -33 52 58.8 & 0.5 &    -            &  14.8 $\pm$ 1.3 & \\
 21 & 17 58 49.58 & -33 57 14.0 & 0.6 &    -            &  14.1 $\pm$ 1.6 & 2mass \\
 22 & 17 58 12.44 & -33 43 56.6 & 0.9 &    -            &  14.0 $\pm$ 1.7 & \\
 23 & 17 58 14.22 & -33 49 08.7 & 0.8 &    -            &  11.7 $\pm$ 1.2 & \\
 24 & 17 58 43.37 & -33 58 17.8 & 1.0 &    -            &  11.6 $\pm$ 1.5 & \\
 25 & 17 59 08.17 & -33 46 47.7 & 0.7 &    -            &   9.5 $\pm$ 1.3 & \\
 26 & 17 58 46.92 & -33 48 44.7 & 1.0 &    -            &   8.0 $\pm$ 1.1 & \\
 27 & 17 58 22.29 & -33 47 43.7 & 1.0 &    -            &   7.9 $\pm$ 1.0 & \\
\enddata

\vspace{-12pt}\tablecomments{Table~\ref{xmmsources} includes for each
source: the source number; RA and DEC -- the position of the source in
J2000 coordinates; Error - the statistical error on the source position
in arcseconds, note that this does not include errors in the overall
astrometry; Flux A -- the source flux for observation A in units of
$10^{-14} \, \rm erg \, cm^{-2} \, s^{-1}$ in the 0.3--10~keV band
calculated assuming a power law spectrum with photon index of 1.5 and
corrected for the Galactic absorption column density of $3.1 \times
10^{21} \rm \, cm^{-2}$; Flux B -- the same for observation B; 
Counterparts - indicates counterparts in \citet{angelini03} (AW2003), 
the 2Mass catalog (2mass), or the WGA catalog (WGA).}  
\end{deluxetable}

\begin{figure*}[tb]
\centerline{\psfig{figure=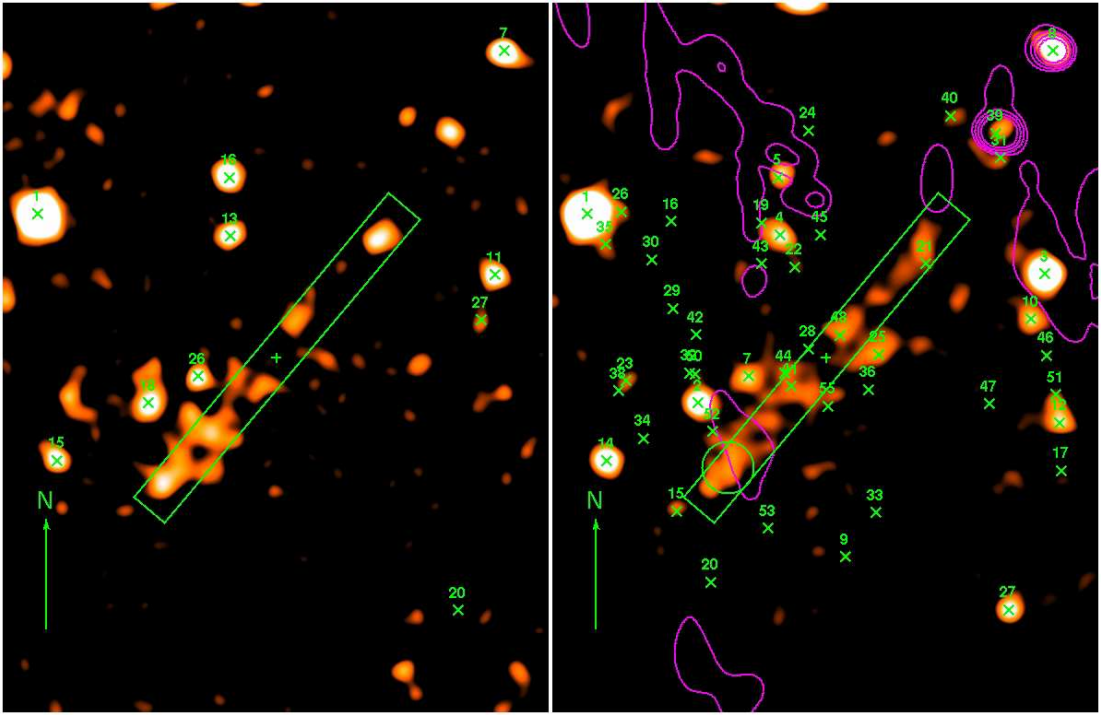,width=5.75in,angle=0} }
\caption{XMM-Newton images of 4U 1755--33.  The image on the left is
observation A (8 Mar 2001) on the right is observation B (18 Sep 2004).
The green cross indicates the position of 4U 1755--33.  The green
rectangle encloses the jet and has a size of $44\arcsec$ by
$430\arcsec$.  On the left image, the green X's mark the positions of
XMM-Newton point sources reported in Table~\ref{xmmsources}.  On the
right image, the green X's mark the positions of Chandra point sources
reported in Table~\ref{chandrasources}.  The green arrow indicates
North and is $2\arcmin$ long.  Radio contours are superimposed on the
image on the right in magenta.  The radio contours represent the flux
density at 13~cm at levels of 0.3, 0.6, 1.2, and 2.4~mJy.  Also, on the
right image, the green circle indicates the extraction region used to
find the spectrum of the jet bright spot described in the text.}
\label{bothobs} \end{figure*}

\section{XMM-Newton Observations and Analysis}

We observed 4U 1755--33 using XMM-Newton beginning on 18 Sep 2004  for
45.9~ks (Observation identifier 0203750101).  We refer to this
observation as ``B''.  We also analyzed an archival observation
obtained by \citet{angelini03} beginning on 8 Mar 2001 with a duration
of 19.4~ks (Observation identifier 0032940101). We refer to this
observation as ``A''. Our observation was designed to have the same
pointing and instrument modes as the earlier observation in order to
facilitate comparison. The EPIC detectors were operated in full frame
mode so that a large field could be imaged.

We reduced the data using the standard procedures described in the 
XMM-Newton User's Handbook and the XMM-Newton ABC Guide.  We used
\textsl{SAS} version 6.1.0.   Calibration files were obtained using the
online version of {\it cifbuild} available at XMM-Newton Science
Operations Centre home page (http://xmm.vilspa.esa.es) on 11 Jan 2005. 
There were no obvious background flares in observation A and only a few
small flares of less than 15\% of the total count rate in observation
B, so we decided not to filter the events on count rate as is often
necessary to remove background flares.  In our analysis, we used only
the EPIC MOS detectors because there are sizable chip gaps in the PN in
the region covered by the jet.  We extracted single, double, and triple
events (PATTERN $\leq$ 12) for the MOS data and filtered according
standard procedures to eliminate events in bad pixels or rows or
frames, in flickering pixels, identified as cosmic-rays, outside the
pulse height thresholds, near the CCD boundary, or outside the nominal
field of view (specifically, we required FLAG equal to \#XMMEA\_EM).

\subsection{Point sources}

We created images in sky coordinates in the 0.3-10~keV band and used
{\it edetect\_chain} to search for sources in the two MOS images for
each observation.  Our primary interest in source detection is the
alignment of the two images, so we considered only sources with very
high significance (likelihood parameter $\ge$ 50, equivalent to a
single trial significance of $17\sigma$) because these sources give the
most accurate positions.  A list of high significance sources is
presented in Table~\ref{xmmsources}.  Many sources of lower
significance are present in the images.  The fluxes are for the
0.3-10~keV band and calculated from the count rate in this band
assuming a power-law spectrum with a photon index of 1.5 and
interstellar absorption with $N_H = 3.1 \times 10^{21} \rm \, cm^{-2}$
equal to the hydrogen column density along the line of sight in the
Milky Way.

We searched for spatial coincidences between the source list from
observation B and the catalog of 2mass point sources.  The astrometric
accuracy of sources in the 2mass catalog is $0.2\arcsec$.  We
considered only sources with J magnitudes brighter than 12.5 to limit
the source density.  We find 1290 sources within $10\arcmin$ of 4U
1755-33 for a source density of 0.0011~arcsec$^{-2}$.  There are 4
matches between X-ray and 2mass sources within $0.9\arcsec$ and one
additional match with an offset of $1.1\arcsec$.  Given the source
density quoted above, the probability of a chance coincidence within
$0.9\arcsec$ between a 2mass source and a given X-ray source is 0.3\%. 
We conclude that the absolute astrometry of the XMM image is good to
within $1.0\arcsec$.  We aligned observation A to observation B by
matching 11 sources detected in both observations.  The needed shift
was $0.15\arcsec$ in RA and $1.3\arcsec$ in DEC.  After the shift, the
average magnitude of offset in source positions between the two
observations is $1.4\arcsec$ which is about twice the typical
statistical uncertainty in the positions as calculated by the SAS
routine {\it emldetect}.  We take $1.4\arcsec$ as an estimate of the
typical uncertainty in the relative XMM-Newton source positions between
the two observations.  Adding in the uncertainty in the absolute
astrometry, we estimate the typical total uncertainty in the XMM Newton
source positions as $1.7\arcsec$.

\subsection{Jet morphology}

Images of the region around 4U 1755--33 are shown in
Fig.~\ref{bothobs}.  The images are for the 0.3--10~keV band.  Each
image is the summed MOS1+MOS2 data in the 0.3--10~keV band and has been
smoothed with a gaussian filter with $\sigma = 8\arcsec$.  The jet
extends to the NW (up and to the right) and SE (down and to the left)
from 4U 1755--33.  The positions of point sources reported in
Table~\ref{xmmsources} are marked on the image.  Note that the image
shown covers only a portion of the field of view, so several sources
listed in the Tables are not present on the image.

\begin{figure}[tb]
\centerline{\psfig{figure=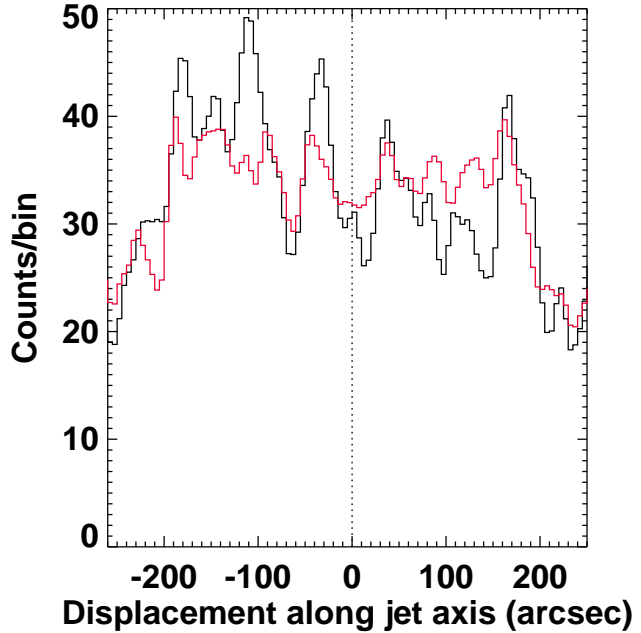,width=3.25in,angle=0} }
\caption{Profile of counts along the jet.  The black line is
observation A and the red line is for observation B.  The counts for
observation A were multiplied by a factor a 2.51 to compensate for the
shorter observation time of that observation.  The dotted line marks
the position of 4U 1755-33.} \label{profile} \end{figure}

Fig.~\ref{profile} shows the profiles of counts along the jet axis. 
The numbers of counts for observation A were multiplied by a factor of
2.51 to compensate for the shorter exposure time.  Each profile
represents the  counts in the 0.5-10 keV band integrated in bins which
extend $2\arcsec$ along the jet axis and have a full width of
$44\arcsec$ perpendicular to the jet axis (the width is the same as the
rectangular region shown in Fig.~\ref{bothobs}).  The profiles were
smoothed with a gaussian with $\sigma = 6\arcsec$.  

The morphology of the jet appears similar in the two observations.
There appears to be some correspondence between knots found in the two
observations, but extraction of knot motions is ambiguous because the
morphologies of the individual knots have changed.  With this caveat,
we consider the morphology of the jet close to 4U 1755--33.  Assuming
that the black hole candidate (BHC) is no longer feeding the jet, one
might expect the gap around the BHC to widen if the jets are moving
away from the BHC.  The knot to the SW closest to the BHC (at a
displacement of $-30\arcsec$ in Fig.~\ref{profile}) does appear to have
shifted away from the BHC by $12\arcsec$ if we compare the peaks of the
two profiles in Fig.~\ref{profile}.  This corresponds to a speed of
$\sim 0.2c$ for a distance of 4~kpc ($\sim 0.4c$ for 9~kpc).  However,
on the NE side of the jet, the closest knot appears to have remained
stationary.  We stress that any conclusions regarding physical motion
of the jets is ambiguous.  In particular, it is unclear if these
changes are really secular or merely random changes of flux over time,
perhaps associated with interactions of the jet with the ISM.

\begin{figure}[tb]
\centerline{\psfig{figure=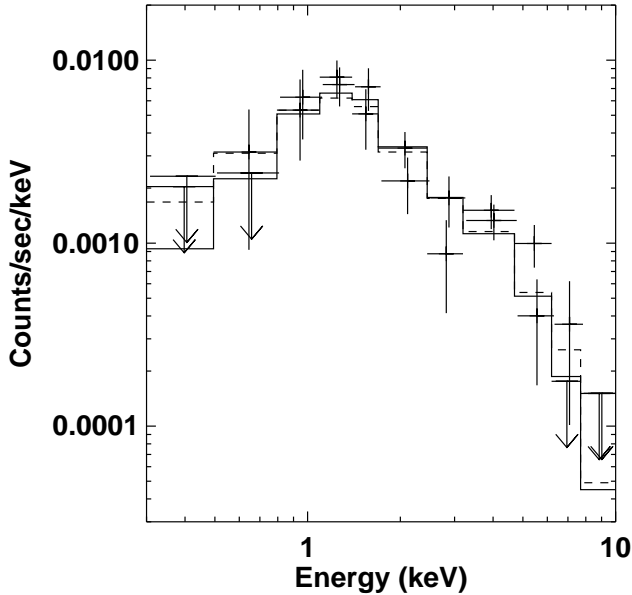,width=3.25in,angle=0} } \caption{X-ray
spectrum of jet emission for observation A.  The data from the two MOS
units are shown.  The energy bins for MOS2 are artifically shifted up
by 2\% for clarity.  The fit of a power-law model with absorption with
$N_H = 3.1 \times 10^{21} \rm \, cm^{-2}$ is shown as a solid line. 
The fit of a mekal model with the same absorption is shown as a dashed
line.} \label{spec} \end{figure}

\begin{deluxetable}{lllccc}
\tabletypesize{\scriptsize}
\tablecaption{X-ray spectral fits to jet emission
  \label{specfits}}
\tablewidth{0pt}
\tablehead{
  \colhead{Model} & \colhead{$N_H$} & \colhead{$\Gamma/kT$} & $\chi^2$/DoF
                  & \colhead{Flux} & \colhead{Unabsorbed Flux} }
\startdata
\multicolumn{6}{l}{Observation A: 8-Mar-2001} \\
 Power-law &  3.1                & $1.7 \pm 0.3$ & 19.3/20 & 2.0 & 2.5 \\
 Power-law & $2.4_{-1.3}^{+2.1}$ & $1.6 \pm 0.4$ & 18.8/19 & 2.1 & 2.5 \\
 MEKAL     &  3.1                & $> 3.7$       & 22.5/20 & 1.9 & 2.4 \\
 MEKAL     & $1.4_{-1.3}^{+2.4}$ & $> 4.5$       & 20.3/19 & 2.1 & 2.3 \\
 \hline
\multicolumn{6}{l}{Observation B: 18-Sep-2004} \\
 Power-law &  3.1                & $1.5 \pm 0.3$ & 13.9/20 & 1.6 & 1.9 \\
 Power-law & $1.9_{-1.3}^{+1.9}$ & $1.4 \pm 0.4$ & 12.5/19 & 1.7 & 1.9 \\
 MEKAL     &  3.1                & $> 5.2$       & 13.8/20 & 1.6 & 1.9 \\
 MEKAL     & $1.8_{-1.3}^{+1.7}$ & $> 6.1$       & 12.2/19 & 1.7 & 1.8 \\
\enddata

\vspace{-12pt}\tablecomments{Table~\ref{specfits} includes for each
spectral model: Model - the model name (note all models include
photoelectric absorption); $N_H$ - the column density for photoelectric
absorption in units of $10^{21} \rm cm^{-2}$; $\Gamma/kT$ - the photon
index $\Gamma$ for power-law models or the temperature $kT$ in keV for
MEKAL models; $\chi^2$/DoF - the $\chi^2$ and the number of degrees of
freedom; Flux -- the source flux in units of $10^{-13} \, \rm erg \,
cm^{-2} \, s^{-1}$ in the 0.5--10~keV band; Unabsorbed Flux - the
source  flux in units of $10^{-13} \, \rm erg \, cm^{-2} \, s^{-1}$ in
the 0.5--10~keV band corrected for interstellar absorption.} 
\end{deluxetable}

\subsection{Jet spectrum and evolution}

We extracted spectra of the jet emission for each observation using the
rectangular regions shown in Fig.~\ref{bothobs}.  Background spectra
were extracted from a region with identical size and orientation
displaced to a source-free region to the Southwest of the jet.  We used
both MOS cameras for each observation and fit the two MOS spectra
simultaneously using \textsl{XSPEC} 11.3.1.

We found adequate fits to the spectra with a power-law subject to
interstellar absorption and with a MEKAL thermal plasma emission model
with solar abundances, again, subject to interstellar absorption, see
Fig.~\ref{spec}.  The fit results are presented in
Table~\ref{specfits}.  We performed fits both with the interstellar
absorption hydrogen column density, $N_H$, fixed to the value found
along the line of sight in H{\sc i} maps \citep{dickey90} and also
allowing $N_H$ to vary.  We used the absorption model of
\citet{wilms00} which is \textsl{tbabs} in \textsl{XSPEC}.  We
calculated the errors on the parameters using $\Delta \chi^2 = 2.71$
corresponding to 90\% confidence for one free parameter of interest for
the fits with $N_H$ fixed, and $\Delta \chi^2 = 4.61$ corresponding to
90\% confidence for two free parameters of interest for the fits with
$N_H$ allowed to vary.  In the table, we report both the observed flux
in the 0.5-10~keV band and the flux corrected for the effects of
interstellar absorption.

Both the power-law and the MEKAL thermal plasma model provide adequate
fits to the spectra.  Allowing $N_H$ to be a free parameter widens the
allowed ranges of the fit parameters.  The preferred values for $N_H$,
when free, tend to be below the column density along the line of sight
toward the source, but include the line of sight density within the
error interval.

No line emission is apparent.  The upper bounds on line emission depend
on the assumed line parameters.  For line parameters similar to those
found for SS433 \citep{watson86}, we find 90\% confidence upper bounds
on the equivalent widths of Fe-K emission lines of 0.6-2~keV.  These
are in the range of the equivalent width measured for the inner jets of
SS433 \citep{migliari02} and are not constraining.

We also fit a spectrum for a region enclosing a bright spot toward the
SE end of the jet for observation B.  The region is shown in
Fig.~\ref{bothobs} and has radii of $28\arcsec$.  This radius was
selected to allow comparison with the radio beam, discussed below.  The
region was located to maximize the X-ray flux within that radius.  We
find an adequate fit with an absorbed power-law model with the
absorption column density fixed to $N_H = 3.1 \times 10^{21} \rm
cm^{-2}$.  We find a photon index $\Gamma = 1.78 \pm 0.52$, a flux of
$2.8 \times 10^{-14} \, \rm erg \, cm^{-2} \, s^{-1}$ in the
0.5--10~keV band, and a flux corrected for absorption of  $3.6 \times
10^{-14} \, \rm erg \, cm^{-2} \, s^{-1}$.

From our two observations, we can make a crude estimate of the decay
time of the jet.  To minimize the dependence on the fitted spectral
model, we re-fitted the data using a power-law model with the photon
index fixed to 1.73 (the best fit value for observation A) with
interstellar absorption with $N_H = 3.1 \times 10^{21} \rm \,
cm^{-2}$.  Adequate fits were found in all cases.  We repeated the same
procedure using a MEKAL model with fixed parameters and find that the
fluxes differ by less than 5\%. The uncertainty in the flux measurement
is dominated by the uncertainty in the background subtraction.  We find
fluxes of $(20.0 \pm 2.2) \times 10^{-14} \, \rm erg \, cm^{-2} \,
s^{-1}$ in the 0.5--10~keV band  for observation A and $(14.6 \pm 1.4)
\times 10^{-13} \, \rm erg \, cm^{-2} \, s^{-1}$ for observation B. 
For an exponential decay, the $1/e$ folding time is $11_{-4}^{+24}$
years.  We also considered the evolution of the Northern and Southern
halves of the jet independently.  For the Southern half of the jet, the
flux is $(12.5 \pm 1.7) \times 10^{-14} \, \rm erg \, cm^{-2} \,
s^{-1}$ in the 0.5--10~keV band  for observation A and $(7.8 \pm 1.0)
\times 10^{-13} \, \rm erg \, cm^{-2} \, s^{-1}$ for observation B. 
The $1/e$ folding time of is $7.6_{-1.3}^{+2.2}$ years.  The flux from
the Northern part of the jet in observation B is consistent within the
uncertainties with that in observation A.  Therefore, we can place only
a lower bound on the decay time of the Northern jet of 8~years.

\section{Radio Observations and Analysis}

On 27 April 2004, we obtained continuum radio observations of
4U~1755$-$33 using the Australia Telescope Compact Array (ATCA) located
in Narrabri, New South Wales, Australia.  This is 153 days before our
XMM-Newton observations, but still reasonably contemporaneous given the
long life-time of the jet.  The ATCA synthesis telescope consists of
five 22~m antennas, which can be positioned along an east-west track
with a short north-south spur, and a sixth antenna at a fixed
location.  The ATCA uses orthogonal polarized feeds and records full
Stokes parameters.  We observed with the ATCA for 10 hours at 1384~MHz
(21.7~cm) and 2368~MHz (12.7~cm) simultaneously, with the array in the
EW367 compact configuration which has baselines ranging from 46 to
4408~m.  Because an interferometer acts as a filter for low spatial
scales, the maximum angular size that can be imaged with the ATCA in
this configuration is of the order of $5\arcmin$ at 2368 MHz
($8\arcmin$ at 1384 MHz).  Since the X-ray jets in 4U~1755$-$33 are
diffuse hot spots, rather than a continuous smooth structure, radio
imaging of the jets is possible with this ATCA configuration.  The
amplitude and band-pass calibrator was PKS~1934$-$638, and the
antenna's gain and phase calibration, as well as the polarization
leakage, were derived from regular observations of the nearby
($6\arcdeg$ away) calibrator PMN~1729$-$37.  The editing, calibration,
Fourier transformation, deconvolution, and image analysis were
performed using the MIRIAD software package \citep{sault98}.

An image of the radio emission at 2.4~GHz (13~cm) is shown as contours
in the right panel of Fig.~\ref{bothobs}.  Strong radio sources were
detected in the northwest region of the field, but there is no
significant detection of emission over the region of the X-ray jet. 
The rms sensitivity is 0.2~mJy/beam.  There is a slight hint of
emission toward the SE end of the jet, but the flux density is always
below 0.6~mJy and does not constitute a detection.  The image at
1.4~GHz also shows no evidence for a significant detection of emission
over the area of the jet.

\begin{deluxetable}{rlllrrl}
\tabletypesize{\scriptsize}
\tablecaption{Chandra X-ray sources near 4U 1755-33
  \label{chandrasources}}
\tablewidth{0pt}
\tablehead{ \colhead{No.} & \colhead{RA} & \colhead{DEC} & \colhead{Error} & 
  \colhead{Flux 2003}  & \colhead{Flux 2004}  &
  \colhead{Counterparts} }
\startdata
  1 & 17 59 00.859 & -33 45 48.67 & 0.1 &                  & 168   $\pm$  7   & [AW2003] 1, 2mass \\
  2 & 17 58 51.181 & -33 49 13.96 & 0.1 &  57   $\pm$  6   &  43   $\pm$  4   & XMM 18 \\
  3 & 17 58 21.041 & -33 46 53.66 & 0.1 &  64   $\pm$  6   & 169   $\pm$  9   & XMM 11 \\
  4 & 17 58 44.051 & -33 46 12.00 & 0.1 &  34   $\pm$  5   &  47   $\pm$  4   & XMM 13 \\
  5 & 17 58 44.259 & -33 45 09.68 & 0.1 &  47   $\pm$  5   &  33   $\pm$  3   & XMM 16 \\
  6 & 17 57 58.976 & -33 46 25.94 & 0.1 &                  & 217   $\pm$ 12   & XMM 2 \\
  7 & 17 58 46.824 & -33 48 45.02 & 0.1 &  26   $\pm$  4   &  19   $\pm$  2   & XMM 26 \\
  8 & 17 58 20.343 & -33 42 51.82 & 0.1 & 171   $\pm$ 16   &  98   $\pm$  7   & XMM 7 \\
  9 & 17 58 38.416 & -33 52 00.99 & 0.3 &  10   $\pm$  3   &  19   $\pm$  2   &  \\
 10 & 17 58 22.234 & -33 47 42.75 & 0.2 &  41   $\pm$  5   &  26   $\pm$  3   & XMM 27 \\
 11 & 17 58 12.455 & -33 43 56.47 & 0.2 &                  &  44   $\pm$  6   & XMM 22 \\
 12 & 17 58 19.725 & -33 49 35.24 & 0.2 &                  &  21   $\pm$  3   &  \\
 13 & 17 58 38.734 & -33 57 00.24 & 0.1 &  57   $\pm$  9   &                  &  \\
 14 & 17 58 59.146 & -33 50 16.97 & 0.2 &  22   $\pm$  4   &                  &  \\
 15 & 17 58 53.104 & -33 51 11.25 & 0.2 &  11   $\pm$  3   &   7.0 $\pm$  1.5 &  \\
 16 & 17 58 53.508 & -33 45 56.84 & 0.2 &                  &   9.9 $\pm$  1.8 &  \\
 17 & 17 58 19.579 & -33 50 27.36 & 0.2 &                  &  14   $\pm$  2   &  \\
 18 & 17 58 05.137 & -33 46 07.22 & 0.2 &                  &  26   $\pm$  4   &  \\
 19 & 17 58 45.698 & -33 45 58.70 & 0.2 &                  &   8.4 $\pm$  1.7 &  \\
 20 & 17 58 50.079 & -33 52 28.71 & 0.2 &                  &   6.7 $\pm$  1.5 &  \\
 21 & 17 58 31.416 & -33 46 42.77 & 0.3 &  12   $\pm$  3   &                  &  \\
 22 & 17 58 42.817 & -33 46 46.27 & 0.1 &                  &   5.6 $\pm$  1.3 &  \\
 23 & 17 58 57.504 & -33 48 49.89 & 0.2 &                  &   5.3 $\pm$  1.3 &  \\
 24 & 17 58 41.580 & -33 44 18.61 & 0.2 &  13   $\pm$  4   &                  &  \\
 25 & 17 58 35.470 & -33 48 21.62 & 0.2 &   5.8 $\pm$  1.9 &   6.4 $\pm$  1.4 &  \\
 26 & 17 58 57.877 & -33 45 46.42 & 0.1 &  15   $\pm$  3   &   5.2 $\pm$  1.4 &  XMM 22 \\
 27 & 17 58 24.178 & -33 52 58.22 & 0.1 &  13   $\pm$  4   &                  &  \\
 28 & 17 58 41.560 & -33 48 15.43 & 0.2 &                  &   4.3 $\pm$  1.2 &  \\
 29 & 17 58 53.396 & -33 47 31.43 & 0.1 &   8   $\pm$  2   &   4.3 $\pm$  1.2 &  \\
 30 & 17 58 55.258 & -33 46 38.68 & 0.3 &                  &   3.7 $\pm$  1.1 &  2mass \\
 31 & 17 58 24.932 & -33 44 47.96 & 0.2 &                  &  14   $\pm$  3   &  \\
 32 & 17 58 51.943 & -33 48 41.59 & 0.1 &                  &   3.7 $\pm$  1.1 &  \\
 33 & 17 58 35.714 & -33 51 13.17 & 0.2 &                  &   4.6 $\pm$  1.3 &  \\
 34 & 17 58 55.982 & -33 49 52.40 & 0.2 &                  &   4.5 $\pm$  1.2 &  \\
 35 & 17 58 59.228 & -33 46 22.15 & 0.2 &                  &   3.6 $\pm$  1.2 &  \\
 36 & 17 58 36.358 & -33 48 59.51 & 0.2 &   5.7 $\pm$  1.9 &   4.0 $\pm$  1.2 &  2mass \\
 37 & 17 57 53.117 & -33 47 01.93 & 0.2 &                  &   8   $\pm$  7   &  \\
 38 & 17 58 58.107 & -33 49 00.99 & 0.2 &                  &   2.9 $\pm$  1.0 &  \\
 39 & 17 58 25.308 & -33 44 21.14 & 0.1 &  12   $\pm$  4   &  20   $\pm$  5   &  \\
 40 & 17 58 29.228 & -33 44 02.49 & 0.2 &   8   $\pm$  3   &  10   $\pm$  2   &  \\
 41 & 17 58 43.075 & -33 48 56.07 & 0.3 &                  &   3.1 $\pm$  1.0 &  \\
 42 & 17 58 51.382 & -33 47 59.83 & 0.3 &   8   $\pm$  2   &                  &  \\
 43 & 17 58 45.690 & -33 46 42.96 & 0.1 &                  &   5.1 $\pm$  1.2 &  \\
 44 & 17 58 43.694 & -33 48 42.10 & 0.3 &                  &   2.9 $\pm$  1.0 &  \\
 45 & 17 58 40.548 & -33 46 11.34 & 0.2 &                  &   5.2 $\pm$  1.4 &  \\
 46 & 17 58 20.885 & -33 48 22.09 & 0.3 &                  &   6.0 $\pm$  1.6 &  \\
 47 & 17 58 25.897 & -33 49 14.88 & 0.4 &                  &   4.8 $\pm$  1.5 &  \\
 48 & 17 58 38.894 & -33 48 00.68 & 0.3 &                  &   3.3 $\pm$  1.0 &  \\
 49 & 17 58 49.783 & -33 58 21.79 & 0.2 &  27   $\pm$  7   &                  &  \\
 50 & 17 58 51.420 & -33 48 42.47 & 0.3 &                  &   3.1 $\pm$  1.0 &  2mass \\
 51 & 17 58 20.112 & -33 49 04.92 & 0.3 &  19   $\pm$  5   &   6   $\pm$  2   &  \\
 52 & 17 58 49.906 & -33 49 44.76 & 0.2 &                  &   2.8 $\pm$  0.9 &  \\
 53 & 17 58 45.112 & -33 51 29.75 & 0.3 &                  &   3.1 $\pm$  1.0 &  \\
 54 & 17 58 42.143 & -33 55 58.54 & 0.2 &   9   $\pm$  3   &                  &  \\
 55 & 17 58 39.876 & -33 49 17.69 & 0.6 &   4.8 $\pm$  1.8 &                  &  \\
 56 & 17 58 11.545 & -33 45 02.64 & 0.4 &                  &   6   $\pm$  3   &  \\
\enddata

\vspace{-12pt}\tablecomments{Table~\ref{chandrasources} includes for
each source: the source number; RA and DEC -- the position of the
source in J2000 coordinates; Error - the statistical error on the
source position in arcseconds, note that this does not include errors
in the overall astrometry; Flux 2003 -- the source flux on 25 Sept 2003
in units of $10^{-15} \, \rm erg \, cm^{-2} \, s^{-1}$ in the
0.3--10~keV band calculated assuming a power law spectrum with photon
index of 1.5 and corrected for the Galactic absorption column density
of $3.1 \times 10^{21} \rm \, cm^{-2}$; Flux 2005 -- the source flux on
25 June 2004;  Counterparts - indicates counterparts in
\citet{angelini03} (AW2003),  the 2Mass catalog (2mass), or the XMM
sources in Table~\ref{xmmsources} (XMM).}   \end{deluxetable}

\section{Chandra Observations and Analysis}

To determine if the features found in the XMM-Newton X-ray image are
point sources or truly diffuse emission, we analyzed two Chandra
observations obtained of the same field.  The observations are a 22~ks
exposure obtained on 25 Sept 2003 (ObsID 3510, PI S.\ Murray) and a
45~ks exposure obtained on 25 June 2004 (ObsID 4586, PI L.\ Angelini).
The 2003 observation is the one analyzed by \citet{park05}.

The Chandra data were subjected to the usual data processing and event
screening and analyzed using the {\it CIAO} version 3.1 data analysis
package, e.g.\ Kaaret (2005).  We constructed images using all valid
events on the S2 and S3 chips for ObsID 3510 and all events on the S3
and S4 chips for ObsID 4586.  We computed exposure maps for each image
for a power-law spectrum with photon index of 1.5 absorbed by a column
density of $N_{H} = 3.1 \times 10^{21} \rm \, cm^{-2}$.  We used the
{\it wavdetect} tool to search for X-ray sources.  The sources with
detection significance of $4\sigma$ or higher are listed in
Table~\ref{chandrasources}. 

We aligned the Chandra observations with sources in the 2mass point
source catalog, as done for the XMM-Newton image above.  We restricted
the sources to have J magnitudes brighter than 12.5 to limit the source
density.  After applying a shift of $0.3\arcsec$ to the Chandra
astrometry, there are 4 matches within $0.2\arcsec$ between Chandra
sources in ObsID 4586 and 2mass sources.  We conclude that our
corrected Chandra astrometry for ObsID 4586 is good to within
$0.2\arcsec$.  We aligned ObsID 3510 to ObsID 4586 using several X-ray
sources present in both images.

Summing the flux from all of the sources within the jet, as defined by
the region shown in Fig.~\ref{bothobs}, we find that discrete point
sources contribute less than 10\% of the total jet flux in each
observation.   Hence, we confirm the results of \citet{park05} that the
X-ray jet from 4U 1755--33 is truly diffuse.  We note that there are
Chandra point source counterparts to the two radio sources beyond the
jet to the northwest.  Therefore, these two radio sources are not
associated with the diffuse X-ray emission of the jet.

To investigate the flux from 4U 1755--33 itself, we extracted counts
from a circular region with a radius of $2\arcsec$ at the position of
the optical counterpart.  We find a total of 3 counts in the 0.3--8~keV
band in the two images.   Using a $15\arcsec$ radius region, we
estimated the background in the source extraction region to be 3.0
counts.  Therefore, we detect zero net counts from 4U 1755--33. 
Allowing for a Poisson distribution of counts, a 95\% confidence upper
limit on the number of source counts is 3.0.  For a power-law spectrum
with photon index of 1.5 absorbed by a column density of $N_{H} = 3.1
\times 10^{21} \rm \, cm^{-2}$, this corresponds to an upper limit on
the source flux of $5 \times 10^{-16} \rm \, erg \, cm^{-2} \, s^{-1}$.
The upper limit on the unabsorbed flux is $6 \times 10^{-16} \rm \, erg
\, cm^{-2} \, s^{-1}$.  For distances of 4--9~kpc, this corresponds to
a limit on the luminosity of $1-6 \times 10^{30} \rm \, erg \, s^{-1}$.
This is similar to the quiescent luminosities of other short orbital
period black hole X-ray binaries such as XTE J1118+480 and A 0620-00
\citep{corbel05b}.

\section{Discussion}

There is now known to exist a broad range of jets from stellar-mass
X-ray binaries.  Persistent compact jets with lengths of tens of A.U.\
are produced in the low/hard X-ray spectral state.  Impulsive jets
produced at state transitions have been detected on lengths scales from
hundreds of A.U.\ out to parsecs.  Stationary lobes have been found in
the radio at separations of several parsecs for sources such as
1E1740.7-2942 and GRS 1758-258 and in the X-ray with separations up to
70~pc from SS 433 \citep{watson83}.  

The jet size of 3--8~pc in 4U 1755--33 is larger than the transient
large-scale moving jets of XTE J1550-564 \citep{corbel02} and
H~1743$-$322 \citep{corbel05}, but is similar to the total size of the
stationary radio lobes of 1E1740.7-2942 and GRS 1758-258
\citep{mirabel99}.  The latter two sources are persistent X-ray
emitters, similar in properties to 4U 1755--33 while it was X-ray
bright.  This may suggest that 4U 1755--33 represents the formation of
a large-scale, nearly stationary jet.  

The central source in 4U 1755--33 has turned off and we now see the
decay of the jet over a time scale of 10--40 years or perhaps longer
for the Northern part of the jet.  While we can place only a lower
bound of 23 years on the time over which the central source was active,
and the jet was being energized, our detection of the decay of the jet
suggests that the time scale for the energization of the jet is similar
to or shorter than the decay time.  We note that we detect flux decay
only for the southern jet.  The change in morphology is also stronger
for the inner regions of the southern jet.  This may suggest that the
southern jet is the approaching and the northern jet is receding. 
Continued monitoring of 4U 1755--33 may reveal motion and decay of the
northern jet.

A key question is the nature of the jet emission.  We consider three
possible emission mechanisms: thermal bremsstrahlung, inverse-Compton,
and synchrotron.  

If the X-ray emission is thermal bremsstrahlung, then the total mass
and energy of the jet can be estimated from its observed luminosity,
temperature, and volume.  For the volume, we assume that the jet
occupies a roughly cylindrical volume with a diameter perpendicular to
the direction of motion of $0.2\arcmin$ and a linear dimension of
$6\arcmin$ multiplied by a filling factor of 0.2.  The volume is then
$2 \times 10^{54} \rm \, cm^{3}$ for an assumed distance of 4~kpc.  We
took the flux from the X-ray spectral fits, and set $kT = 5 \rm \, keV$
which is the minimum temperature consistent with the fits.  We find
that the density of the jet material is $4 \rm \, cm^{-3}$, the total
energy of the jet is $2 \times 10^{47} \rm \, erg$ and the total mass
in the jet is $1 \times 10^{31} \rm \, g$.  The cooling time of the gas
would then be $6 \times 10^{6} \rm \, yr$ which is much longer than the
observed decay time.  Also, if the jet were fed by the outflow from a
mass accretion rate of $10^{19} \rm \, g/s$ corresponding the Eddington
rate for a $10 M_{\odot}$ compact object, at least $10^{4}$ years would
be required to accumulate the needed mass.  This is much longer than
the observed decay time scale and we conclude that the jet emission is
unlikely to be thermal bremsstrahlung.

The best candidate source of seed photons for inverse-Compton
scattering is the interstellar radiation field (ISRF).  The ISRF varies
strongly with Galactocentric radius and height above the Galactic
plane.  On the sky, 4U 1755-33 is rather close to the Galactic center
and it could be physically close to the Galactic center, although the
low optical extinction suggests that the source is actually nearer than
the Galactic bulge.  We assume that 4U 1755-33 is at a distance of
8.5~kpc, close to the Galactic center, which maximizes the ISRF energy
density.  We adopt an energy density of 10~eV~cm$^{-3}$, equal to the
maximum found anywhere in the Milky Way \citep{strong00} and assume,
for simplicity, that all of the radiation is in the dominant component
of ISRF near 1~$\mu$m \citep{mathis83}.  To produce X-rays in the
0.3-10~keV, electrons with energies from $\sim$5 to $\sim$100~MeV are
required.  For an assumed X-ray spectral index $\alpha = -0.5$, defined
as $S(\nu) \propto \nu^{\alpha}$, the total energy in relativistic
electrons required is $\sim 10^{50} \rm \, erg$.  A $10 M_{\sun}$ black
hole producing energy at the Eddington rate with all of the energy
going to perfectly efficient acceleration of electrons in the desired
energy range would require $\sim 2000$ years to power the jet.  This is
much longer than the observed decay time scale.  Therefore, it appears
unlikely that the jet emits via inverse-Compton radiation.

If X-ray emission is synchrotron, then synchrotron radio emission
should be expected.  To determine if our upper limits on the radio flux
are consistent with synchrotron emission, we consider the emission from
the brightest X-ray spot along the jet.  The X-ray spectrum was
extracted from a circle with a radius of $28\arcsec$.  This is slightly
larger than the radio beam size at 2.4~GHz, so we can compare the X-ray
flux density in the regions to the radio upper limit per beam, which we
take as 3 times the rms noise level of 0.2~mJy.  Comparing the X-ray
and radio flux density, we derive a lower limit on the radio/X-ray
spectral index $\alpha > -0.65$, i.e. the spectrum must be flatter than
$\alpha = -0.65$.   Therefore, our radio observations may simply be not
sensitive enough to detect the radio synchrotron emission.  The limit
on the spectral index implies that the exponent, $p$, of the electron
energy distribution, $N(E) \propto E^{-p}$, must be $p < 2.3$.  This is
within the range that can be produced by relativistic shocks.   The
spectral index is similar to that measured for the large-scale jets of
the black hole candidate X-ray transient XTE J1550-564 of $\alpha =
-0.660 \pm 0.005$ \citep{corbel02} and consistent with the index of
$\alpha = -0.45$ measured for the eastern lobe of SS 433
\citep{safiharb99}.

To investigate the energetics of a jet radiating via synchrotron
emission, we calculate the equipartition magnetic field.  Using a
spectral index $\alpha = -0.6$ which is consistent with the bound
above, a lower frequency cutoff of 2.4~GHz, and the same assumption for
distance and volume as for the thermal bremsstrahlung case, we find a
magnetic field of 36~$\mu$G.  The electrons producing the X-ray
emission must have Lorentz factors up to $6 \times 10^7$, corresponding
to energies of up to 60 TeV.  The radiative lifetime of these electrons
is of order 320~years.  The minimum total energy required is  $2 \times
10^{44} \rm \, erg$ and the number of electrons needed to produce the
observed radiation is $1.2 \times 10^{46}$.  Assuming that the jet is
composed of normal matter (i.e.\ roughly one proton per electron), then
the required mass is $2 \times 10^{22} \rm \, g$.  These numbers are
relatively insensitive to distance.  For a distance of 9~kpc, the
magnetic field decreases to 29~$\mu$G, the required energy increases to
$1.5 \times 10^{45} \rm \, erg$, and the required mass increases to
$1.3 \times 10^{23} \rm \, g$.  For the larger distance, the energy
required corresponds to 2 weeks accumulation at the Eddington rate for
a 10~$M_{\sun}$ black hole and the mass could be accumulated in less
than one day.  Thus, the energy and material required for a synchrotron
emitting jet could easily have been accumulated over the 20 year active
phase of the X-ray source.  Using the decay rate calculated for the
entire jet and the energy estimate from the synchrotron equipartition
calculation, the energy loss rate is then $\sim 4 \times 10^{35} \rm \,
erg \, s^{-1}$ for a 4~kpc distance.  This is about 1\% of the
Eddington luminosity for a 10~$M_{\sun}$ black hole and, therefore, the
jet could have been energized by the conversion of a few percent of the
energy into relativistic electrons.

The available data on the jet of 4U 1755--33 are consistent with the
X-rays being synchrotron emission.  Synchrotron emission also is the
most favorable mechanism in terms of the required mass and energy.
However, significantly deeper radio observations are required to test
if the predicted synchrotron radio emission is produced.

\acknowledgments

We thank an anonymous referee for useful comments and the XMM-Newton
team for successfully executing the observation.  PK acknowledges
partial support from NASA grant NNG05GA08G and a faculty scholar award
from the University of Iowa.  JAT acknowledges partial support from
NASA grant NNG04GQ05G.

{}

\end{document}